\documentclass[aps,twocolumn,amsmath,amssymb,prb,showpacs]{revtex4-1}
\usepackage{amsfonts, hyperref, times}
\usepackage{graphicx}

\begin{document}

\title{Electric signature of magnetic domain-wall dynamics}

\author{Y.~Liu}
\author{O.~A.~Tretiakov}
\author{Ar.~Abanov}

\affiliation{
             Department of Physics \& Astronomy,
	    Texas A\&M University,
            College Station, Texas 77843-4242, USA
}

\date{August 16, 2011}

\begin{abstract}
Current-induced domain-wall dynamics is studied in a thin
ferromagnetic nanowire. The domain-wall dynamics is described by
simple equations with four parameters.  We propose a procedure to
unambiguously determine these parameters by all-electric measurements
of the time-dependent voltage induced by the domain-wall motion. We
provide an analytical expression for the time variation of this
voltage.  Furthermore, we show that the measurement of the proposed
effects is within reach of current experimental techniques.
\end{abstract}


\maketitle

\textit{Introduction.} Recently, applications for future memory and
logic devices, as well as important fundamental physics questions,
have stimulated a number of experimental \cite{Yamaguchi04,
  Thomas2006, Thomas2007, Meier07, Rhensius10, IlgazPRL10,
  Krivorotov10, BeachPRL09} and theoretical \cite{Beach:review08,
  YanAPL10, YanEPL10} studies of the current-driven domain wall (DW)
dynamics in ferromagnetic nanowires. It has been shown that DWs can be
moved by a current either parallel \cite{Yamaguchi04, Thomas2006,
  Thomas2007, Meier07, Rhensius10, IlgazPRL10} or perpendicular to the
wire. \cite{Krivorotov10, YanAPL10, YanEPL10} In some of the
experiments short current pulses were employed to depin a DW from
pinning sites. \cite{Thomas2006, Thomas2007, IlgazPRL10}  Furthermore,
the topological electromotive force induced by DW dynamics in a vortex
DW has been studied both experimentally and theoretically.
\cite{BeachPRL09, Yang2010}

A conventional experimental method to study the DW dynamics in
nanowires is to measure the average DW velocity using Kerr
polarimetry, \cite{Beach05} x-ray microscopy, \cite{Meier07} or
electron microscopy. \cite{Rhensius10, Klaui:images05} These types of
experiments require a complicated setup which is separate from the one
needed for the DW manipulation.  This situation is neither ideal for
studies of DW dynamics nor for further technological advances.

In this Letter we propose a way to use the same experimental setup for
both current DW manipulation and simultaneous measurements of DW
dynamics.  Our main results are that the time-dependent voltage
induced by the DW motion \cite{Tserkovnyak2008, Duine09} can be used
to fully and comprehensively determine the effective parameters of the
DW dynamics.  This proposal follows from the fundamental properties of
the current-induced DW motion, namely: (i) Applied DC current (above
critical value) produces voltage with AC components.  (ii) Applied AC
current induces phase shifted AC voltage.  The magnitude of the
proposed effects is calculated to be within current experimental
resolution. 

Similar techniques have already shown promise in magnetic field driven
DW systems. \cite{Singh10} This method should make it more feasible to
utilize DW dynamics for device applications. Furthermore, the proposed
systematic approach can be used to compare the extracted
phenomenological parameters of the DW dynamics for a system described
by arbitrary underlying Hamiltonian to those of microscopic theories.

\textit{Model.}  The dynamics of the magnetization $\mathbf{S}$ in a
quasi-one-dimensional wire is described by Landau-Lifshitz-Gilbert
(LLG) equation with current $j$, \cite{Zhang04, Thiaville05}
\begin{equation}
  \dot{\mathbf{S}}=-\mathbf{S}\times\mathbf{H}_{e}
-j\partial_z \mathbf{S}
+\beta j\mathbf{S}\times\partial_z \mathbf{S}
+\alpha\mathbf{S}\times\dot{\mathbf{S}},
\label{eq:LLG}
\end{equation}
where $\mathbf{H}_{e}=-\delta\mathcal{H}/\delta\mathbf{S}$ is the
effective magnetic field given by the Hamiltonian $\mathcal{H}$ of the
system, $\mathbf{S}=\mathbf{M}/|M|$ is a unit magnetization vector,
$\alpha$ is the Gilbert damping constant, $\beta$ is the non-adiabatic
spin torque constant, $\partial_z \equiv \partial/\partial z$ where
$\mathbf{\hat z}$ is along the wire, and the time is measured in units
of the gyromagnetic ratio $\gamma_0 =g|e|/(2mc)$.  DWs in a
ferromagnetic wire can be modeled by a spin Hamiltonian $\mathcal{H}$
which contains exchange, spin-orbit, \footnote{It results in
  crystalline anisotropy, Dzyaloshinskii-Moriya interaction, etc.}
and dipolar interactions. In a thin wire, the latter can be
approximated by two anisotropies: a strong anisotropy along the wire
($\lambda$) and a weak anisotropy transverse to it ($K$). In realistic
systems $\alpha, \beta\ll 1$ and $K\ll \lambda$.

\begin{figure}
\includegraphics[width=1\columnwidth]{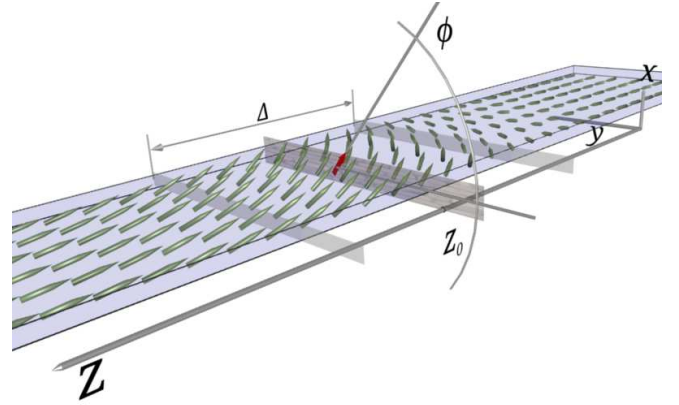}
\caption{(Color online) A moving head-to-head domain wall of width
  $\Delta$. The DW is centered at $z_0$ and is tilted by an angle
  $\phi$.}
\label{DWpicture}
\end{figure}

In a thin wire, a lowest-energy magnetization configuration (at $j=0$)
is uniformly ordered along the $z$ or $-z$ direction. A static DW is
the next low-energy configuration with the boundary conditions $S_z
(\pm \infty ) =\pm 1$ or $S_z (\pm \infty ) =\mp 1$. DWs can be
injected in the wire using different techniques.  A sketch of a wire
with a DW of width $\Delta$, determined by the Hamiltonian parameters,
is depicted in Fig.~\ref{DWpicture}. 

For small enough applied currents, it can be shown that the DW in a
thin wire is a rigid spin texture \cite{Klaui:images05} and its
dynamics can be described in terms of only two collective
coordinates. \cite{Tatara04, Tretiakov_DMI} These coordinates
correspond to the two softest modes of the DW motion: the DW position
along the wire, $z_0$, and the rotation angle $\phi$ of the
magnetization in the DW around the wire axis, see
Fig.~\ref{DWpicture}.  It has been shown \cite{Tretiakov:losses,
  Tretiakov_DMI} that the equations of motion for the DW in a thin
ferromagnetic wire are model independent and can very generally be
written in the form
\begin{eqnarray}
\label{z}
&&\dot{z}_0=Aj+B[j-j_c\sin(2\phi)],\\
\label{phi}
&&\dot{\phi}=C[j-j_c \sin(2\phi)].
\end{eqnarray}
Here all current nonlinearities are neglected, since the large
currents leading to observable nonlinear effects would burn the
nanowire.  For a dc current below the critical value $j_c$, i.e.,
$j<j_c$, Eq.~\eqref{phi} implies that the DW tilts from the transverse
anisotropy plane by the angle that satisfies $\sin(2\phi)=j/j_c$
around the wire axis and then moves along the wire with a constant
velocity $Aj$. For $j>j_c$, the DW constantly rotates while moving.

The coefficients $A$, $B$, $C$ and the critical current $j_c$ are the
parameters that fully describe the DW dynamics. They can be calculated
microscopically for certain toy models, \cite{Tretiakov_DMI} but in
general they vary for different wires and depend on the temperature
and nanofabrication details. Therefore, in this Letter we propose a
way to determine these coefficients by model-independent measurements
of an induced ac voltage directly from an experiment suitable for
all-electric DW manipulation. As we show below, this ac voltage can be
induced by applied dc currents and by certain time-dependent current
pulses with parameters similar to those achieved in recent
experiments. \cite{KlauiPRL05, KlauiAPL10}

Microscopically the dynamics parameters can be obtained in the
following way. The energy of a static DW, $E_0 (z_{0},\phi) = \int
\mathcal{H} [\mathbf{S}_0(z,z_{0},\phi)] dz$, where $\mathbf{S}_0$ is
a solution of a static LLG with $K=0$, in general depends on both
$z_{0}$ and $\phi$. However, assuming that the wire is translationally
invariant (pinning can be neglected), $E_0$ would not depend on the DW
position $z_0$ and therefore $\partial_{z_0} E_0=0$. The only
contribution to $E_0$ that depends on the angle $\phi$ comes from the
small anisotropy in the transverse plane, $E_0 (\phi) = -\kappa
\cos(2\phi)$. \cite{Tretiakov_DMI} \footnote{For the Hamiltonian of
  Ref.~\onlinecite{Tretiakov_DMI} this constant is $\kappa =\pi
  K\Gamma \Delta^2/\sinh(\pi \Gamma \Delta)$ which reduces to $\kappa
  =K \Delta$ for $\Gamma\Delta\ll 1$.} This allows us to find the
coefficients in Eqs.~\eqref{z} and~\eqref{phi} in terms of the
parameters of the LLG~\eqref{eq:LLG}. \cite{Tretiakov_DMI, future} Up
to first order in $\alpha$ and $\beta$ they are
\begin{eqnarray}
\label{A}&&A=\frac{{\tilde \beta}}{{\tilde \alpha}}, \qquad 
B=\frac{{\tilde \alpha}-{\tilde \beta}}{{\tilde \alpha}}(1
+{\tilde \alpha} a_{z\phi}),\\
\label{C}&&C= ({\tilde \alpha}-{\tilde \beta}) a_{zz},\qquad
j_c = \frac{{\tilde \alpha}}{{\tilde \alpha}-{\tilde \beta}}
\kappa,
\end{eqnarray}
where ${\tilde \alpha}= \alpha \mathcal{D}$, ${\tilde \beta}= \beta
\mathcal{D}$, $\mathcal{D} = \sqrt{a_{zz}a_{\phi\phi} -a_{z\phi}^2}$,
$a_{zz}=\frac{1}{2} \int dz (\partial_z \mathbf{S}_0)^2$,
$a_{\phi\phi}=\frac{1}{2} \int dz (\partial_{\phi} \mathbf{S}_0)^2$,
and $a_{z\phi}= \frac{1}{2} \int dz \partial_z \mathbf{S}_0\cdot
\partial_{\phi} \mathbf{S}_0$.  Equations~\eqref{A} and \eqref{C} are
consistent \footnote{For the Hamiltonian used in
  Ref.~\onlinecite{Tretiakov_DMI}, $\mathcal{D} =1$,
  $a_{z\phi}=\Gamma\Delta$, $a_{zz}=(1+\Gamma^2\Delta^2)/\Delta$, and
  $\Gamma=D/J$, where $D$ and $J$ are, respectively, the
  Dzyaloshinskii-Moriya and exchange interaction constants.} with the
expressions for $A$, $B$, $C$, and $j_c$ found in
Ref.~\onlinecite{Tretiakov_DMI}.

We now outline the method to find $A$, $B$, $C$, and $j_c$ directly
from all-electric measurements. It is based on measuring the ac
voltage $V$ induced by a moving DW.  To find $V$ one has to know the
time evolution of the total energy (per unit area of the wire's
cross-section) in the system,
\begin{eqnarray}
\dot{E}&=&\int dz\frac{\delta \mathcal{H}}{\delta \mathbf{S}}
\cdot\dot{\mathbf{S}}(z). 
\end{eqnarray}
In general, DW energy has two contributions: the power supplied by an
electric current and a negative contribution due to dissipation in the
wire.  Using the general solution of the LLG, Eq.~\eqref{eq:LLG}, one
can obtain the derivative of the energy as \cite{Tretiakov_DMI,
  future}
\begin{eqnarray}
\label{E_dot}
\dot{E} = 2[\beta a_{zz} \dot{z}_{0} +(1 -\beta  a_{z\phi }) \dot{\phi }]j
- \alpha \int dz \dot{\mathbf{S}}_0^2 .
\end{eqnarray}
The last term on the right-hand side of Eq.~\eqref{E_dot} describes
the dissipation and is therefore always nonpositive. Meanwhile, the
first term is proportional to the current density $j$ and gives the power
$Vj$ supplied by the current. With the help of Eqs.~\eqref{A}--\eqref{C}
and adopting the approximation $\mathcal{D}\simeq 1$ of
Ref.~\onlinecite{Tretiakov_DMI} we obtain the expression for the
induced DW voltage \footnote{Since in majority of materials both
  $\alpha \ll 1$ and $a_{z\phi}=\Gamma\Delta \ll 1$, we can be safely
  neglect $\alpha\Gamma\Delta$ compared to 1 in Eq.~\eqref{voltage}.},
\begin{equation}
\label{voltage}
V=\frac{A^2C}{B}j +C(1+A)[j-j_c \sin (2\phi)].
\end{equation}
Note that Eq.~\eqref{voltage} gives the contribution to the voltage
due to DW motion. This contribution is in addition to the usual Ohmic
one.  The voltage $V$ in Eq.~\eqref{voltage} is measured in units of
$Pg\mu_B/(e\gamma_0)$ and the current density is measured in units
$2eM/(Pg\mu_B)$, where $P$ is the current polarization. We emphasize
that unlike in the previously studied cases, \cite{BeachPRL09,
  Yang2010} this voltage is not caused by the motion of topological
defects (vortices) transverse to the wire.

\textit{Measurement of coefficients $A$, $B$, $C$, and $j_c$.}  In
order to find coefficients $A$, $B$, and $C$, we propose three
independent measurements of the voltage induced by a moving DW.
Although there are various factors affecting the nanowire resistance,
the contributions from most of them are independent of DW motion and
therefore give only a constant component of the resistance. To
characterize the DW dynamics, one has to concentrate only on the
resistance variations in time. Our estimates show that the amplitude
of voltage oscillations due to DW motion is of the order of $10^{-7}$
V and therefore experimentally measurable.

Equation~\eqref{voltage} implies that the voltage of the DW can give
all the necessary information about DW dynamics. Namely, one can
obtain $C$ by measuring the voltage changing with time and
parameters $A$ and $B$ by measuring the amplitude of the voltage
oscillations.

\begin{figure}
\includegraphics[width=1\columnwidth]{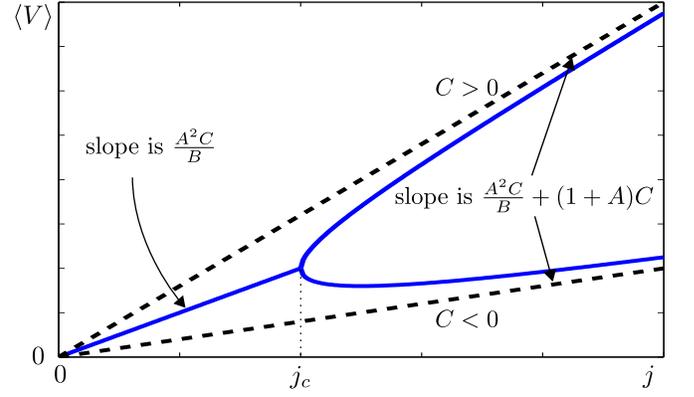}
\caption{(Color online) Dependence of average voltage $\langle V
  \rangle$ on dc current $j$ for $C>0$ and $C<0$, respectively, see
  Eq.~\eqref{voltage}. The slope at $j<j_c$ gives $\frac{A^2C}{B}$,
  whereas at $j\gg j_c$ it gives $\frac{A^2C}{B}+(1+A)C$.}
\label{a_V_j}
\end{figure}

\textit{Slopes measurement.} In Refs.~\onlinecite{Tretiakov:losses,
  Tretiakov:JAP} it was proposed to obtain $A$, $B$, and $j_c$ by
measuring the drift velocity of the DW, $\langle \dot{z}_0
\rangle$. It is important to note that Eq.~\eqref{voltage} has the
same form as Eq.~\eqref{z}. Thus, instead of measuring the drift
velocity, which requires a more complicated experimental setup, we
propose to perform all-electric measurements. Namely, to measure the
average voltage of DW, $\langle V \rangle$, as a function of dc
current. From Eq.~\eqref{voltage} one can see that $\langle V
\rangle=\frac{A^2C}{B}j$ for $j<j_c$, whereas $\langle V
\rangle=\frac{A^2C}{B}j +(1+A)C\sqrt{j^2-j_c^2}$ for $j>j_c$, see
Fig.~\ref{a_V_j}. The critical current is determined by the end of the
region linear in $j$ for small currents. The measurement of slope
$k_1$ at $j<j_c$, and slope $k_2$ at $j\gg j_c$ gives the two
independent quantities:
\begin{equation}
\label{k1}
k_1=\frac{A^2C}{B},\qquad k_2-k_1=(1+A)C.
\end{equation}
Instead of measuring voltage average for dc current, one can apply a
linearly increasing time-dependent current $j(t)=q t$ below the
critical value $j_c$.  At sufficiently small $q$ the voltage will also
be linear in time, $V(t)\approx \frac{A^2C}{B} q t$. By measuring this
voltage one can find
\begin{equation}
\frac{V(t)}{j(t)}=\frac{A^2C}{B}.
\end{equation}
Once $C$ is determined, Eqs.~\eqref{k1} give $A$ and $B$.  The
drawback of this measurement is that it might be hard to disentangle
$k_1$ and $k_2$ from the Ohmic contribution. However $k_2-k_1$ is free
from the Ohmic resistance of the wire.

\begin{figure}
\includegraphics[width=1\columnwidth]{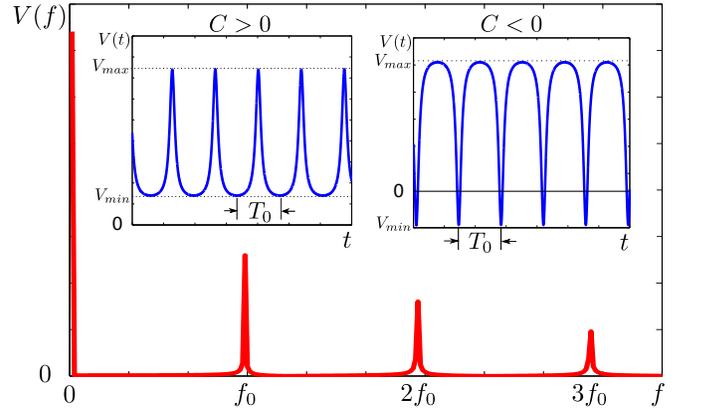}
\caption{(Color online) Fourier transform of the voltage $V$ as a
  function of frequency $f$ at the dc current $1.1 j_c$. The insets show
  $V$ as a function of time $t$ for $C>0$ given by $\alpha=0.02$ and
  $\beta=0.01$; and for $C<0$ given by $\alpha=0.01$ and $\beta=0.02$.
  The voltage period is $T_0=1/f_0$. In the inset for $C<0$, the
  voltage varies between $V_{max}=0.041 j_c/\Delta$ and
  $V_{min}=-0.019 j_c/\Delta$. }
\label{fft}
\end{figure}

In order to find $C$, the most intuitive approach is to input a dc
current slightly above $j_c$. Then the voltage induced by the moving
DW will oscillate with the period of the double angle $\phi$, see the
insets of Fig.~\ref{fft}. The half-width of the peak (dip) for $C>0$
($C<0$) is given by $\arccos(j_c/j)/(|C|\sqrt{j^2-j_c^2})$. The
measurement of the voltage oscillations period $T_0$ (which we
estimate to be $\sim 10^{-7}$ -- $10^{-6}$ s) determines $C$ at a
given $j$:
\begin{equation}
|C|=\frac{1}{T_0}\int_0^\pi \frac{d\phi}{j-j_c\sin(2\phi)}
=\frac{\pi}{T_0\sqrt{j^2-j_c^2}}.
\end{equation}
For $j-j_c\ll j_c$, the period diverges but the half-width $\sim
1/(Cj_c)$ stays finite.  To obtain the period $T_0$, one can perform the
Fourier transform of $V(t)$ to find the frequency $f_0= 1/T_0$, see
Fig.~\ref{fft}.

To determine coefficient $A$ in the same experiment, one can measure
$\Delta V=V_{max}-V_{min}=2(1+A)|C|j_c$, see insets of
Fig.~\ref{fft}. Then
\begin{equation}
A=\frac{\Delta V}{2|C|j_c}-1.
\end{equation}
Note that $\Delta V=2(k_2 -k_1)j_c$ and therefore this experiment can
also provide a crosscheck with the aforementioned measurement of the
slopes.

\begin{figure}
\includegraphics[width=1\columnwidth]{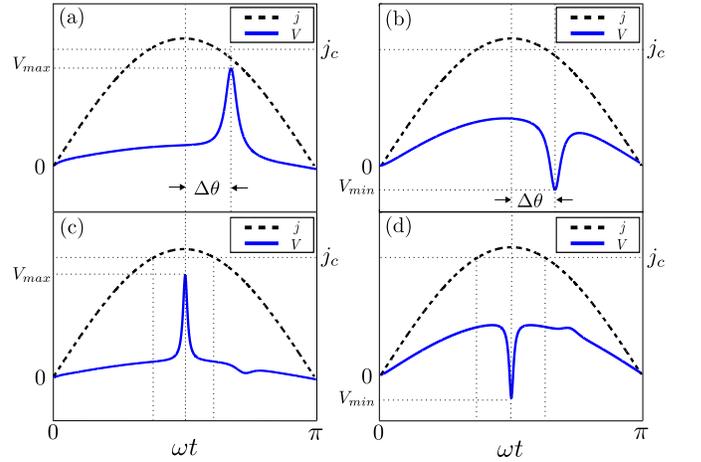}
\caption{(Color online) Input current $j$ (dashed line) and measured
  voltage $V$ (solid line) as functions of time $t$. (a) and (b) show
  the phase delay $\Delta\theta$ between the current maximum and
  voltage extremum for $C>0$ and $C<0$, respectively.  (c) and (d)
  depict $V(t)$ at $\Delta\theta=0$ for the same $C>0$ and $C<0$,
  respectively.}
\label{pulse}
\end{figure}

\textit{Phase shift experiment.} Another method to measure the
coefficient $C$ is by applying an ac current $j=j_{0}\sin \omega t$
with $j_{0}>j_c$, which has only a short time interval where $j>j_c$,
so that there is only one period of voltage within the period of
$j(t)$.  One can measure the phase delay, $\Delta \theta$, between the
current maximum and voltage extremum \footnote{Our simulations show
  that the initial phase of angle $\phi$ does not affect the phase
  delay, since the time it takes for the current to increase from $0$
  to $j_c$ is long enough to adjust the initial angle to the one
  corresponding to $\sin(2\phi)\approx j/j_c$.} (see Fig.~\ref{pulse}).
Next, one fixes the amplitude $j_0$ and tunes the frequency $\omega$
until $\Delta \theta=0$.  In this case, for $j_0-j_c\ll j_c$, we can
use half of the time interval for which the current pulse is above
$j_c$ to approximate the period of $\phi$ by dc current $j_0$ as
\begin{eqnarray}
\label{pulse_app1}
\frac{1}{2\omega}\left( \pi -2\arcsin \frac{j_c}{j_0} \right)
\approx\frac{\pi}{|C|\sqrt{j_0^2-j_c^2}}.
\end{eqnarray}
For $j_0-j_c\ll j_c$, Eq.~\eqref{pulse_app1} can be further simplified
to give
\begin{eqnarray}
\label{pulse_app2}
|C|\approx\frac{\pi \omega}{2(j_0 -j_c)}.
\end{eqnarray}
In other words, when $\omega \approx C(j_0-j_c)$ which corresponds
roughly to $\omega \sim 10^{7}$ Hz, the current pulse covers only one
period of voltage.  Our simulations show that the
expression~\eqref{pulse_app2} works sufficiently well for $j_0
\lesssim 1.3 j_c$. The sign of $C$ is determined by the extremum of
the measured voltage: $C>0$ if $V$ has the minimum and $C<0$ if $V$
has the maximum.

Our simulations show (Fig.~\ref{pulse}) that in addition to the large
peak (dip) of voltage there is a smaller one with the opposite
curvature. This is because when $j(t)$ reaches $j_c$, the angle $\phi$
has not yet rotated to the angle corresponding to $\sin(2\phi_0)=1$
due to the cumulative phase delay between current and voltage.

\begin{figure}
\includegraphics[width=1\columnwidth]{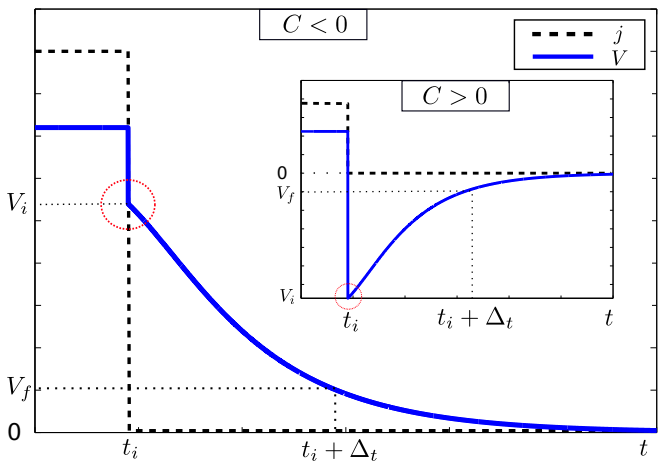}
\caption{(Color online) Voltage (solid line) evolution after the
  current (dashed line) is turned off at time $t_i$ for $C>0$ given by
  $\alpha=0.02$ and $\beta=0.01$. Inset: the same dependencies for
  $C<0$ given by $\alpha=0.01$ and $\beta=0.02$. The measurement of
  $V_f$ is performed at $t_i +\Delta_t$. The region encircled by the
  dotted line cannot be described within our approach but it is too
  small to effect our results.}
\label{long_tail}
\end{figure}

\textit{Abrupt current pulse experiment.} It is also possible to
measure the coefficient $C$ for currents below the critical value
$j_c$. The constant $|C|j_c$ determines the internal time scale of the
DW motion.  After one switches the subcritical current off at time
$t_i$, the voltage asymptotically decays as $\exp(-2|C|j_c t)$, see
Fig.~\ref{long_tail}.  To measure the decay of $V(t)$ with time, one
inputs a dc current below $j_c$, then measures voltage $V_i$
immediately after turning off the current at $t_i$, and then later
measures voltage $V_f$ at time $t_i + \Delta_t$.  We note that right
after turning off the current, there is a short time period when the
DW dynamics cannot be described by Eqs.~\eqref{z} and \eqref{phi}. It
corresponds to the dynamics of fast degrees of freedom. This process
has a characteristic time $\sim 10^{-11}s$ which is typically much
smaller than the voltage decay time $\sim 10^{-8}$ s. Thus we can
safely assume that the rotation angle $\phi$ does not change much
during this time interval, and we find
\begin{equation}
|C| \simeq \frac{1}{2\Delta_t j_c}\ln
\frac{2V_i/V_f}{1+\sqrt{1-j^2/j_c^2}},
\end{equation}
which is valid for $V_i/V_f \gg 1$. For example, estimating $V_i/V_f
=10$ we find $|C|\approx 1.17/(\Delta_t j_c)$. The sign of $C$ can be
easily determined by the form of voltage decay (see
Fig.~\ref{long_tail}).

To summarize, we propose several all-electric measurements of the
parameters fully describing domain-wall dynamics in thin ferromagnetic
nanowires. These measurements are based on the voltage induced by a
moving DW in response to certain current pulses. Our proposal opens
doors for experiments which are suitable not only for all-electric DW
manipulation but also for the simultaneous measurement of the DW
dynamics.  These findings give a more reliable and straightforward
experimental method to determine the DW dynamics parameters, which can
then be compared to microscopic theories. The procedure we described
works for a given temperature regime. It may also be used to
investigate the temperature dependence of the effective
parameters. Future work will include accounting for pinning effects,
which brake translational invariance in the wires. \cite{future2}

We thank I.~V.~Roshchin, J.~Sinova, and E.~K.~Vehstedt for valuable
discussions.  This work was supported by the NSF Grant No. 0757992 and
Welch Foundation (A-1678).

\bibliography{magnetizationDynamics}

\end{document}